\documentstyle[12pt,fullpage]{article}	

\renewcommand{\baselinestretch}{1.05}

\newenvironment{tabAlgorithm}[2]{
\setcounter{algorithmLine}{1}
\samepage
\begin{tabbing}
999\=\kill
#1 \ \ --- \ \ \parbox{4in}{\it #2}
}{
\end{tabbing}
}
\newcounter{algorithmLine}
\newcommand{\algline}{\\\thealgorithmLine\hfil\>\stepcounter{algorithmLine}}

\newcommand{\qed}{\vspace{.1em}\noindent\fbox{\rule{
0em}{.1em}\rule{.1em}{0em}}\vspace{1em}}

\newenvironment{proof}{

\noindent{\bf Proof:}\ }{
\hfill \qed

}

\newcommand{\R}{\Re}		

\newtheorem{myLemma}{Lemma}

\title{
A Primal-Dual Parallel Approximation Technique \\
Applied to Weighted Set and Vertex Cover}

\author{
\large
{\it Samir Khuller}\ \thanks{
Computer Science Department and Institute for Advanced Computer Studies
(UMIACS), University of Maryland, College Park, MD 20742.
Part of the work of this author was done while this author was with UMIACS
and
partially supported by NSF grants CCR-8906949, CCR-9111348, and CCR-9103135.
Email: samir@cs.umd.edu.}
\and
{\it Uzi Vishkin}\ \thanks{
Department of Electrical Engineering and Institute for Advanced 
Computer Studies (UMIACS), University of Maryland,
College Park, MD 20742. Also with Tel-Aviv University.
Partially supported by NSF grants CCR-8906949 and CCR-9111348.
Email: vishkin@umiacs.umd.edu.}
\and
{\it Neal Young}\ \thanks{
Institute for Advanced Computer Studies (UMIACS), University of Maryland,
College Park, MD 20742.
Partially supported by NSF grants CCR-8906949 and CCR-9111348.
Email: young@umiacs.umd.edu.}
}
\date{}
\begin{document}
\maketitle

\begin{abstract}
We give an efficient deterministic parallel approximation algorithm
for the minimum-weight vertex- and set-cover problems
and their duals (edge/element packing).
The algorithm is simple and suitable for distributed implementation.
It fits no existing paradigm for fast, efficient parallel algorithms
--- it uses only ``local'' information at each step, 
yet is {\em deterministic}.
(Generally, such algorithms have required randomization.)
The result demonstrates 
that linear-programming primal-dual approximation techniques
can lead to fast, efficient parallel algorithms.
The presentation does not assume knowledge of such techniques.
\end{abstract}



\section{Introduction}
\label{introSec}

The linear-programming primal-dual method
for obtaining sequential algorithms
for exact optimization problems is well studied \cite{chv}.
Primal-dual techniques have also been used
to obtain sequential approximation algorithms
(e.g., for NP-hard problems \cite[etc.]{chv-79,hoch-82}
and for on-line problems \cite{you}).
In this paper, we apply primal-dual techniques to obtain
a deterministic parallel approximation algorithm
for the minimum-weight vertex- and set-cover problems
and their duals, maximum-weight edge and element packing. 

The result demonstrates that linear-programming primal-dual techniques
can lead to fast, efficient parallel algorithms.
The algorithm is natural,
yet fits no existing paradigm for such algorithms,
being unique in that it uses only ``local'' information at each step, 
yet is {\em deterministic}.
Generally, such algorithms have required randomization
(e.g., \cite{II-86,ABI,lu-86}).

Given an $n$-vertex, $m$-edge graph $G=(V,E)$ with vertex weights
and an $\epsilon > 0$,
our algorithm returns a vertex cover of weight
at most $2/(1-\epsilon)$ times the minimum.
It uses $O(\ln ^2 m\,\ln\frac{1}{\epsilon})$ time and $m/\ln ^2 m$ processors
(i.e., $O(m\, \ln\frac{1}{\epsilon})$ operations) on an EREW-PRAM.
More generally, the algorithm finds a set cover of weight
at most $r/(1-\epsilon)$ times the minimum,
using $O(r\,\ln^2 m\,\ln \frac{1}{\epsilon})$ time and $M/\ln^2 m$ processors
(i.e., $O(r\, M\, \ln \frac{1}{\epsilon})$ operations).
Here $m$ is the number of elements,
$M$ is the sum of the set sizes,
and $r$ is the maximum number of sets in which any element occurs.
(For vertex cover, $r=2$.)
In each case, the algorithm also implicitly finds a near-maximal dual solution
(an edge or element {\em packing})
that is also within the corresponding factor of optimal.

The algorithm can be implemented using only integer arithmetic
(see \S \ref{precisionSec}).
If the weights are integers and $1/\epsilon$
is less than the sum of the vertex (resp.~set) weights,
then the weight of the cover is at most 2 (resp.~$r$) times the minimum.

\subsection{Related Work}
The first $r$-approximation algorithm for weighted vertex/set cover
was due to Hochbaum \cite{hoch-82}.
She considered the relaxation of the natural integer linear program
for the problem.
The dual of this program is maximum edge packing.
The so-called complimentary-slackness conditions 
are that a (fractional) cover and a packing are optimal provided
(i) every vertex in the cover has its constraint met in the packing 
and (ii) every edge with non-zero packing weight
has exactly one vertex in the cover.
Hochbaum observed that an optimal packing was necessarily maximal,
that for any maximal packing,
the vertex set formed by the vertices whose packing constraints are met
with equality
form a cover, 
that such a packing and cover satisfy (i),
and that (i) is sufficient to guarantee $r$-approximation
because (ii) is approximately satisfied in that every edge
has at most $r$ vertices in the cover.
Since an optimal dual solution can be found
in polynomial time by solving the linear program,
Hochbaum obtained a polynomial-time algorithm.
Bar-Yehuda and Even \cite{BE-81} observed that 
sequentially raising the edge-packing weights as much as possible
yields a maximal edge packing, thus obtaining a linear-time algorithm.
For our algorithm, we relax (i) further,
insisting only that every vertex in the cover 
{\em nearly} have its constraint met,
and we show how to simultaneously raise {\em many} edge-packing weights
so that the packing quickly becomes {\em nearly} maximal
and the weight of the cover formed by the vertices
that nearly have their packing constraints met
with equality
is within $r/(1-\epsilon)$ of optimal.

In \cite{clar-83}, Clarkson showed that in a restricted class of graphs,
approximation ratios better than 2 could be obtained for vertex cover.
Clarkson gave the first parallel approximation algorithm
--- a relatively complicated randomized algorithm \cite{clar-91}.
According to Motwani's lecture notes on approximation algorithms 
\cite{mot}, which contain a survey of results on vertex cover,
the best approximation ratio known
is $2-\frac{\log\log n}{2\log n}$,
due to Bar-Yehuda and Even \cite{BE-85}
and to Monien and Speckenmeyer \cite{MS}.
In \cite{hoch-83}, Hochbaum gives a $(2-2/k)$-approximation algorithm,
where $k$ is the maximum vertex degree,
and she conjectures that there is no polynomial-time 
$c$-approximation algorithm for any $c < 2$ unless P$=$NP.

Chv\'atal's weighted-set-cover algorithm
guarantees a set cover of weight
at most $\ln \Delta$ times the minimum,
where $\Delta$ is the maximum set size
\cite{chv-79,lov-75,joh-74}.
Berger, Rompel, and Shor 
\cite{BRS-89} give a parallel algorithm
that guarantees a factor of $(1+\epsilon)\ln \Delta$.
Their algorithm uses a linear number of processors
and runs in polylogarithmic time with some restrictions on the weights.

The intuition behind our complexity analysis
relies on a lemma of general interest for parallel graph algorithms
(Lemma \ref{happyCamperLemma}).
The lemma has previously found application
in the analyses of randomized parallel graph algorithms:
Israeli and Itai's maximum-matching algorithm \cite{II-86}
and Alon, Babai and Itai's maximal-independent-set algorithm \cite{ABI}.

In concurrent independent work,
Cohen gives a parallel approximation algorithm
for maximum flow in shallow networks \cite{cohe-92}.
If network flows are viewed as packings of source-to-sink paths,
then maximal packings correspond to blocking flows.
Cohen gives an $\epsilon$-blocking flow algorithm
that is similar in spirit to our algorithm,
although a number of different issues arise.
In a more recent work, Luby and Nisan
give a parallel primal-dual approximation algorithm
for positive linear programming \cite{LN-93}.
Hochbaum's original algorithm can be parallelized
by employing Luby and Nisan's algorithm; 
the resulting algorithm would obtain an approximation ratio
comparable to ours and have an incomparable running time
(growing linearly with $1/\epsilon$, but not with $r$).
Previously, Goldberg {\em et~al.} \cite{GPST-92} 
gave a parallel primal-dual algorithm
to find (exactly) maximum-weight bipartite matchings.
Their algorithm appears to be the first parallel algorithm 
to use primal-dual techniques, but it requires polynomial time.

\subsection{Problem Definitions}
Let $G=(V,E \subseteq 2^V)$ be a given hypergraph 
with vertex weights $w:V\rightarrow \R^+$.
Let $E(v)$ denote the set of edges incident to vertex $v$.
Let $G$ have $m$ edges.
Let $r$, the {\em rank} of $G$, be the maximum size of any edge.
(For an ordinary graph, $r=2$.)
Let $M$, the {\em size} of $G$, be the sum of the edge sizes.
For any real-valued function $f$
and a subset $S$ of its domain,
let $f(S)$ denote $\sum_{x\in S} f(x)$.

\paragraph{Vertex Cover.}
A {\em vertex cover} for $G$ is a subset $C\subseteq V$ of the vertices
such that for each edge $e \in E$, some vertex in $e$ is in $C$.
The {\em (minimum-weight) vertex-cover problem}
is to find a vertex cover with minimum total weight $w(C)$.

\paragraph{Edge Packing.}
An {\em edge packing} is an assignment $p:E\rightarrow\R^+$
of non-negative weights to the edges of the hypergraph
such that the total weight $p(E(v))$ assigned to the edges
incident to any vertex $v$ is at most $w(v)$.
The {\em (maximum-weight) edge-packing problem}
is to find an edge packing maximizing $p(E)$, the {\em weight} of $p$.
The fractional relaxations of the vertex cover 
and edge packing problems are linear programming duals.

\subsection{Related Problems}
\label{relProbSec}
Let ${\cal C}$ be a family of sets
with weights $w:{\cal C}\rightarrow \R^+$.
Let $U$ denote $\bigcup_{S\in\cal C} S$.

\paragraph{Set Cover.}
A {\em set cover} is a subfamily $\cal C'\subseteq C$
such that $\bigcup_{S\in\cal C'} S = U$ ---
in words, every element of $U$ is in some set in the cover.
The {\em (minimum-weight) set-cover problem}
is to find a set cover of minimum total weight $w(\cal C')$.

\paragraph{Element Packing.}
An {\em element packing} is an assignment
of non-negative weights to the elements
such that the total weight assigned to the elements 
of any set $S$ is at most $w(S)$.
The {\em (maximum-weight) element-packing problem}
is to find an element packing maximizing the net weight assigned to elements.

\subsection{Equivalences}
The vertex cover problem in hypergraphs
is equivalent to the minimum-weight set-cover problem as follows.
For each $S \in {\cal C}$ we have a vertex $v_S$ in the hypergraph.
For each element $x \in U$, we have an edge that {\em contains} $v_S$
if and only if $x \in S$.
The number of edges $m$ is the number of elements.
The rank $r$ is the maximum number of sets in which any element occurs.
The size $M$ is the sum of the set sizes.
The dual problems are also equivalent.

\section{Reduction of Vertex Cover to $\epsilon$-Maximal Packing}
\label{reduceSec}
We first reduce our problem
to the problem of finding what we call an {\em $\epsilon$-maximal} packing. 
This reduction generalizes \cite{hoch-82,BE-81},
who considered $\epsilon = 0$.
\begin{myLemma}[Duality]
\label{dualLemma}
Let $C$ be an arbitrary vertex cover and $p$ an arbitrary edge packing.
Then $p(E) \le w(C)$.
\end{myLemma}
\begin{proof}
\[ p(E) = \sum_{e\in E} p(e)
	\le \sum_{e\in E} |e\cap C|\,p(e)
	= \sum_{v\in C} p(E(v)) 
	\le \sum_{v\in C} w(v)
	= w(C).\]
\end{proof}

\begin{myLemma}[Approximate Complimentary Slackness]
\label{maximalLemma}
Let $C$ be a vertex cover and $p$ be a packing such that
$p(E(v)) \ge (1-\epsilon) w(v)$ for every $v\in C$.
Then $(1-\epsilon) w(C) \le r\,p(E)$.
By duality, the weights of $C$ and $p$ are within
a factor of $r/(1-\epsilon)$ from their respective optima.
\end{myLemma}
\begin{proof}
Since $(1-\epsilon)w(v) \le p(E(v))$ for $v\in C$,
\[(1-\epsilon)w(C) 
	= (1-\epsilon)\sum_{v\in C} w(v)
	\le \sum_{v\in C} p(E(v))
	= \sum_{e\in E} |e\cap C|\,p(e)
	\le r\,p(E).\]
\end{proof}
We tighten Lemma \ref{maximalLemma}
slightly when the weights are integers.
\begin{myLemma}
\label{integerLemma}
In Lemma \ref{maximalLemma}, 
if the weights are integers and $\epsilon < 1/w(V)$,
then the weight of $C$
is at most $r$ times the minimum.
\end{myLemma}
\begin{proof}
Let $C^*$ be a minimum-weight cover.  From Lemma \ref{maximalLemma},
$(1-\epsilon) w(C) \le r\, w(C^*)$, so
$w(C) \le \lfloor r\,w(C^*) + \epsilon\,w(C) \rfloor
= r\, w(C^*)$.
\end{proof}

Given a packing $p$,
define $C_p = \{v \in V : p(E(v)) \ge (1-\epsilon)w(v)\}$.
If $C_p$ is a vertex cover,
then we say $p$ is {\em $\epsilon$-maximal}.
Note that $p$ is $0$-maximal if and only if $p$ is maximal.
By Lemma \ref{maximalLemma}, 
if $p$ is $\epsilon$-maximal,
then $C_p$ and $p$ are within a factor of $r/(1-\epsilon)$
from their respective optima.

\section{The Algorithm}
\label{algSec}
We have reduced the problem to finding an $\epsilon$-maximal packing.
The algorithm maintains a packing $p$
and the partial cover
$C_p = \{v \in V: p(E(v)) \ge (1-\epsilon)w(v)\}$.
The algorithm increases the individual $p(e)$'s 
until $p$ is $\epsilon$-maximal and $C_p$ is a cover.
When a vertex $v$ enters $C_p$,
$v$ and the edges containing $v$ are deleted from the hypergraph.
Let $E_p$ denote the set of remaining edges,
let $E_p(v)$ denote the remaining edges incident to vertex $v$,
and let $d_p(v)$ be the degree of $v$ in $G_p=(V,E_p)$.
Define the {\em residual weight $w_p(v)$} of vertex $v$ to be $w(v)-p(E(v))$.

In a single round of the algorithm,
for each remaining edge $e$, $p(e)$ is raised.
To ensure that $p$ remains a packing,
each vertex $v$ limits the increase in each $p(e)$ for $e\ni v$
to at most $w_p(v)/d_p(v)$.
Each $p(e)$ is then increased
as much as possible subject to the limits imposed by all the $v\in e$.
That is, each $p(e)$ is increased by $\min_{v\in e}w_p(v)/d_p(v)$.
The algorithm repeats this basic round
until $p$ converges to an $\epsilon$-maximal packing. 
It then returns $C_p$.
To implement the algorithm we maintain $w_p$ instead of $p$:
\begin{tabAlgorithm}{{\sc Cover}$(G=(E,V), w, \epsilon)$}{
Returns a vertex cover of hypergraph $G$
of weight at most $r/(1-\epsilon)$ times the minimum.}
\algline {\bf for} \= $v\in V$ {\bf par-do} \(w_p(v) \leftarrow w(v)\);
					\(E_p(v) \leftarrow E(v)\);
					\(d_p(v) \leftarrow |E(v)|\)
\algline {\bf while} \= edges remain {\bf do}
\algline   \>   {\bf for} each remaining edge $e$ {\bf par-do}
		 \(\delta(e) \leftarrow 
			\min_{v \in e} w_p(v)/d_p(v)\)
\algline   \>  {\bf for} \= each remaining vertex $v$ {\bf par-do}
\algline   \>  \> \(w_p(v) \leftarrow w_p(v) 
			- \sum_{e\in E_p(v)} \delta(e)\)
\algline   \>  \> {\bf if} \= $w_p(v) \le \epsilon\,w(v)$ {\bf then}
\algline   \>  \> \> \= delete $v$ and incident edges, 
	updating $E_p(\cdot)$ and $d_p(\cdot)$
\algline {\bf return} the set of deleted vertices
\end{tabAlgorithm}

As noted above, the limit on the increase in each $p(e)$
ensures that $p$ remains a packing.
Consequently, the correctness
and the approximation ratio of the algorithm
are established by Lemmas \ref{maximalLemma}
and \ref{integerLemma}.
Using standard techniques \cite{Jaja-92}, 
each iteration of the {\bf while} loop
beginning with $q$ remaining edges can be done
in $O(\ln q)$ time and $O(r\,q)$ operations
on an EREW-PRAM.

\section{Complexity Analysis}
\label{analysisSec}

In this section, we prove our main theorem:
\vspace{0.1in}

\noindent
{\bf Main Theorem}\ \ {\it
The algorithm requires
$O(r\,\ln^2 m\,\ln\frac{1}{\epsilon})$ time and $M/\ln^2 m$ processors,
i.e., $O(r\, M\, \ln\frac{1}{\epsilon})$ operations. 
}
\vspace{0.1in}

We use a potential function argument.
Given a packing $p$, define
\[\phi_p = \sum_{v\in V}d_p(v)\ln \frac{w_p(v)}{\epsilon\, w(v)}.\]
The next lemma shows 
that during an iteration of the {\bf while} loop
$\phi_p$ decreases 
by at least the number of edges 
remaining at the end of the loop.
This is how we show progress.

\begin{myLemma}
\label{hypergraphLemma}
Let $p$ and $p'$, respectively, be the packing before and after
an iteration of the {\bf while} loop.
Then $\phi_p - \phi_{p'} \ge |E_{p'}|$.
\end{myLemma}
\begin{proof}
During the iteration, we say that
a vertex $v$ {\em limits} an incident edge $e \in E_p$
if $v$ determines the minimum in the computation 
of $\min_{v\in e} w_p(v)/d_p(v)$.
For each vertex $v$, let $v$ limit $L(v)$ edges, so that
\(w_{p'}(v) \le w_p(v)(1 - L(v)/ d_p(v)).\)  
Let $V'$ denote the set of vertices that remain after the iteration.
Then
\begin{eqnarray*}
\phi_p - \phi_{p'} & = &
        \sum_{v\in V} \left(d_p(v)\ln\frac{w_p(v)}{\epsilon\,w(v)}
                - d_{p'}(v)\ln\frac{w_{p'}(v)}{\epsilon\,w(v)}\right) \\
& \ge & \sum_{v\in V'} d_p(v)\ln\frac{w_p(v)}{w_{p'}(v)} \\
& \ge & \sum_{v\in V'} -d_p(v)\ln (1-L(v)/d_p(v)) \\
& \ge & \sum_{v\in V'} L(v) \\
& \ge & |E_{p'}|
\end{eqnarray*}
The second-to-last step follows
because $-\ln(1-x) \ge x$.
The last step follows because each of the
edges that remains is limited by some vertex in $V'$.
\end{proof}

\begin{myLemma}
There are at most
$(1+r\ln\frac{1}{\epsilon})(1 + \ln m)$ iterations.
\end{myLemma}
\begin{proof}
Let $p$ and $p'$, respectively, be the packing 
before and after any iteration.
Let $a = r\ln\frac{1}{\epsilon}$.

Clearly $\phi_{p'} \le |E_{p'}|\,a$.
By Lemma \ref{hypergraphLemma}, $\phi_{p'} \le \phi_p - |E_{p'}|$.
Thus, $\phi_{p'} \le \phi_p(1-1/(a+1))$.
Before the first iteration, $\phi_p \le m\,a$.
Inductively, before the $i$th iteration,
\[\phi_{p} 
\le m\,a (1-1/(a+1))^{i-1}
\le m\,a\exp(-(i-1)/(a+1)).\]
The last inequality follows from $e^x \ge 1+x$ for all $x$.
Fixing $i=1+\lceil (a+1)\ln m\rceil$,
we have
$\exp(-(i-1)/(a+1)) \le \exp(-\ln m) = 1/m$,
so before the $i$th iteration, $\phi_{p} \le a $.

During each subsequent iteration, at least one edge remains, 
so $\phi_p$ decreases by at least $1$.
Thus, $\phi_p \le 0$ before an $i+a$th iteration can occur.
\end{proof}

\paragraph{Time.}
As each iteration requires $O(\ln m)$ time,
the above lemma implies that the total time
is $O(r\ln^2 m \ln\frac{1}{\epsilon})$.

\paragraph{Operations.}
Recall that an iteration with $q$ edges
requires $O(r\,q)$ operations.
Consequently, the total number of operations is bounded
by an amount proportional
to $r$ times the sum, over all iterations,
of the number of edges at the beginning of that iteration.

By Lemma \ref{hypergraphLemma}, in a given iteration,
$\phi_p$ decreases by at least the number of edges
remaining at the end of the iteration.
Thus, the sum over all iterations
of the number of edges during the iteration
is at most $m+\phi_p$ for the initial $p$.
This is $m+M\ln\frac{1}{\epsilon}$.
Hence there are $O(r\,M\,\ln\frac{1}{\epsilon})$ operations.

\paragraph{Processors.}
Using standard techniques, the operations can be efficiently scheduled without
increasing the time or the operations
by more than a constant, so by the Work-Time Scheduling Principle
\cite{Jaja-92},
the number of processors required is $M/\ln^2 m$ ---
the work divided by the time.
This establishes the Main Theorem.

\subsection{The Intuition for Ordinary Graphs}
\label{intuitSec}
The potential function analysis, while easy to verify,
hides an interesting combinatorial principle
that gives a good intuitive understanding 
of the algorithm for ordinary graphs.
Recall that,
during a single iteration of the {\bf while} loop,
a vertex $v$ {\em limits} an edge $e$
if $v$ determines the minimum in the calculation of 
\(\min_{v\in e} w_p(v)/d_p(v),\)
and that $L(v)$ denotes the number of edges limited by $v$.
If a vertex $v$ limits at least a third of its incident edges,
then $w_p(v)$ decreases by at least one third its value.
Call such a vertex {\em good}.
(After $O(\ln\frac{1}{\epsilon})$ iterations 
of the {\bf while} loop in which $v$ is good,
$v$ will enter $C_p$.)
In a given iteration, few vertices might be good.
However, at least half of the remaining edges touch good vertices.
This is a consequence of the following lemma:

\begin{myLemma}[\cite{II-86,ABI}]
\label{happyCamperLemma}
Consider a directed graph.
Call a vertex {\em good} if more than
one-third of its incident edges are directed into it.
Then at least half of the edges are directed into good vertices.
\end{myLemma}
\begin{proof}
If a vertex is not good, call it {\em bad}.
The in-degree of any bad vertex is at most half its out-degree,
so the number of edges directed into bad vertices
is at most half the number of edges directed out of bad vertices.
Thus, the number of edges directed into bad vertices
is at most half the number of edges.
Thus, at least half the edges are directed into good vertices.
\end{proof}

To see why the lemma applies, 
imagine directing each remaining edge into a vertex that limits it.
Then the lemma shows that at least half of the
remaining edges touch vertices that are good.
Thus, in a given iteration, at least half the edges
touch vertices whose residual weights decrease 
by more than a factor of $1/3$.
This, intuitively, is why the algorithm makes progress.

This lemma is of independent interest: 
it drives the analyses of the running times 
of Israeli and Itai's randomized maximal matching algorithm \cite{II-86}
and of Alon, Babai, and Itai's
randomized maximal independent set algorithm \cite{ABI}.
Interestingly, the natural generalization 
of the lemma to hypergraphs is not strong enough 
to give an analysis as tight as our potential function analysis.

\subsection{Using Integer Arithmetic}
\label{precisionSec}
If arithmetic precision is an issue,
we can uniformly scale the original (integer) vertex weights
so that the smallest weight is at least $m/\epsilon$,
and then use integer division
(taking the floor) when computing the $w_p(v)/d_p(v)$'s.
Essentially the same analysis carries through.
(If $w(v) \ge m/\epsilon$,
then $w_p(v) \ge m$ while $v$ remains,
so $w_p(v)/d_p(v) \ge 1$,
hence $\lfloor w_p(v)/d_p(v) \rfloor \ge (w_p(v)/d_p(v))/2$,
and the net reduction in a $w_p(v)$ 
during an iteration is at least half
what it would have been without taking the floor.
Thus, the analysis will go through
by doubling the potential function.)

Assuming without loss of generality that $\epsilon \ge 1/(2w(V))$,
if the original weights are $k$-bit integers, 
then the largest weight after scaling is bounded by
\[ m\left\lceil\frac{1}{\epsilon}\right\rceil 2^{k+1} 
\leq 2^{k+2}m\,w(V) 
\leq 2^{2k+3}m\,|V|. \]
Hence the scaled weights are $(2k+3+\log_2 m+\log_2|V|)$-bit integers.
Subsequently all operations involve only integer arithmetic
on smaller, non-negative integers.

\vspace{0.2in}
\noindent
{\bf Acknowledgments:}
We would like to thank Michael Luby and an anonymous referee for pointing 
out connections to the randomized algorithms of \cite{ABI,lu-86}
and to the work by Hochbaum \cite{hoch-82}, respectively.

\end{document}